\documentclass[aps,twocolumn,showpacs,showkeywords,footinbib]{revtex4}
\usepackage{mathrsfs}
\usepackage{mathtools}
\usepackage{txfonts}%
\usepackage[latin9]{inputenc}
\usepackage{amssymb}
\usepackage{amscd}
\usepackage{graphicx}
\usepackage{subfigure}
\usepackage{float}

\begin{document}
\title[Short Title]{Distributed manipulation of two-qubit entanglement with coupled continuous variables}

\author{Li-Tuo Shen}
\author{Rong-Xin Chen}
\author{Huai-Zhi Wu}
\email{huaizhi.wu@fzu.edu.cn}
\author{Zhen-Biao Yang}
\email{zbyang@fzu.edu.cn}


\affiliation{Lab of Quantum Optics, Department of Physics, Fuzhou
University, Fuzhou 350002, China}

\begin{abstract}
We study the dynamics of two qubits separately sent through two
coupled resonators, each initially containing a coherent state
field. We present analytical arguments and numerical calculations
for the qubit-field system under different two-qubit initial states,
photon hopping strengths, and detunings. In far off-resonant regime,
the maximal entanglement of two qubits can be generated with the
initial qubit state in which one qubit is in the excited state and
the other is in the ground state, and the initially maximal
two-qubit entanglement can be frozen and fully revived even for
large mean photon number. When the qubits are both initially in
their excited states or ground states, the qubit-qubit entanglement
birth and death apparently appear in the regime where the photon
hopping strength is close to qubit-field detuning, and its peaks do
not decrease monotonically as the interaction time increases. It is
interesting to observe that when there is photon hopping strength
between two fields, the field-field entanglement can be larger than
one and increases as the initial amplitude of the coherent state
grows. By postselecting the fields both in their coherent states,
the entanglement of two initially unentangled qubits can be largely
improved. Our present setup is fundamental for the distributed
quantum information processing and applicable to different physical
qubit-resonator systems.
\end{abstract}

   \maketitle

\noindent

\section{Introduction}

Manipulation of the entanglement dynamics between the qubit and
light field has become a significantly important issue for both the
fundamental quantum theories and experiments
\cite{RMP-70-101-1998,PRL-92-127902-2004,PRL-91-070402-2003,
PRL-85-2392-2000,PRL-87-037902-2001,OC-176-265-2000,Nature-411-166-2001,Nature-453-1023-2008},
where reliable quantum information processing (QIP) and computation
rely on coherent manipulation of physically realizable systems in
which information is stored and by means of which information is
processed or transmitted. However, only the dynamics in a
qubit-field system where the field evolves in low-dimensional
subspace is analytically easy to handle.

Since the continuous-variable (CV) physical system contains an
infinite-dimensional spectrum of eigenstates and can be efficiently
generated by a classical monochromatic current \cite{QuantumOptics},
the CV system, as a kind of excellent quantum resource that most
resembles a classical electromagnetic field
\cite{RMP-77-513-2005,JPA-40-7821-2007}, has attracted much
attention in many fields of QIP recently, such as quantum transport
\cite{PRA-75-032336-2007,JPB-44-105501-2011}, quantum storage
\cite{PRA-79-062317-2009,JPB-39-5143-2006}, and quantum memory
\cite{PRL-96-080501-2006,PRL-98-140504-2007}.

Based on resonant Jaynes-Cummings (JC) interaction, Lee \emph{et
al.} \cite{PRL-96-080501-2006} demonstrated that an ebit could
reciprocate between two non-local qubits and two separate coherent
states via postselection, in which the CV systems are able to
reliably accumulate more than one ebit when a series of qubit pairs
interact with the CV systems; Y\"{o}na\c{c} and Eberly
\cite{OL-33-270-2008} reported the collapse and revival behaviors of
entanglement of two separate qubits each interacting with a CV
system; Guo \emph{et al.} have shown that coherent-state control and
entanglement transfer between two non-local qubits and two spatially
separated CV systems are possible \cite{PRA-86-052315-2012}.

In general experiment associated with entanglement manipulation,
there may exist two common modulations, i.e., the hopping strength
between the CV systems and the detuning in the JC interaction.
Previous efforts typically concentrate on the entanglement dynamics
between a pair of non-local qubits and the spatially separate CV
systems, where the dynamics is analytically solvable based on
separate JC models. The coupled-cavity system involving two
non-local qubits and the coupled thermal fields has been
investigated under hopping and detuning modulations, which can
exhibit interesting features, such as maximal qubit-qubit
entanglement generation and freezing \cite{JOSAB-29-2379-2012}.
However, the dynamics involving two non-local qubits and the coupled
CV systems has not been extensively investigated due to their
infinite-dimension Hilbert space and complicated mutual-interaction
processes when considering the hopping between the CV systems. To
our knowledge, a convincingly analytical treatment of the dynamics
between two non-local qubits and the coupled CV systems is still
absent, which is relevant with the recent progresses in the arrays
of interacting micro-cavities and their coupling to qubits
\cite{PRA-78-063805-2008,LPR-2-527-2008,
PRL-96-010503-2006,PRA-82-042327-2010,JOSAB-29-2379-2012,Nature-484-195-2012}.

In this paper, we present the numerically exact solution to the
dynamics between two qubits and two coupled coherent-state fields.
Our setup, differing further from the previous setups, where two
sites each involving a qubit and a CV field evolve independently
based on the resonant Jaynes-Cummings interaction
\cite{PRL-96-080501-2006,OL-33-270-2008}, focuses on the modulation
induced by the hopping between the CV fields and the detuning
between the qubit and the local CV field. The entanglement dynamics
between two qubits depends on initial two-qubit states, photon
hopping strengths, and qubit-field detunings. We present analytical
arguments and numerical calculations for the qubit-field system
within far off-resonant regime. In far off-resonant regime, the
maximal entanglement of two qubits can be generated with the initial
qubit state in which one qubit is in the excited state and the other
is in the ground state, while the initially maximal two-qubit
entanglement can be frozen. Particularly, the maximal qubit-qubit
entanglement can be fully revived by tuning the qubit-field detuning
even when the mean photon number of the coherent state field is
large, which is obviously different from the resonant case where the
maximal entanglement can not be fully revival in Ref.
\cite{OL-33-270-2008}. When the qubits are both initially in their
excited states or ground states, the qubit-qubit entanglement
apparently appears in the regime where the strength of photon
hopping is close to qubit-field detuning. It is interesting to
observe that when there is hopping between two CV systems, the
field-field entanglement can be larger than one and increases as the
initial amplitude of the coherent state grows. By postselecting the
fields both in their coherent states, the entanglement of two
initially unentangled qubits can be largely improved. Even under the
far off-resonant condition, the initially unentangled qubits can
become the maximally entangled by measuring the non-maximally
entangled fields both in their coherent states.

The present idea can be generalized to other coupled CV systems,
including the squeezed coherent state and displaced coherent state,
and has the potential application in the distributed QIP, such as
the quantum state transfer and entanglement protection. Our present
results are fundamental and promising for the distributed QIP and
applicable to different physical qubit-resonator systems.

\section{System Hamiltonian}

%

\begin{figure}[H]\label{fig1}
\centering
\includegraphics[width=0.9\columnwidth]{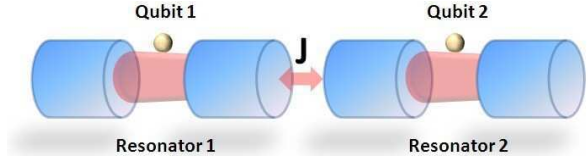}
\caption{(Color online) Schematic of our setup for two non-local
qubits sent through two coupled resonators, each interacting with a
CV system. }\end{figure}

To investigate the entanglement dynamics between qubits and the
coupled continuous variables, let us first consider that two
identical qubits $1$ and $2$ are respectively sent through two
coupled resonators, each initially containing a coherent state field
$|\alpha\rangle_{i}$ ($i$ $=$ $1$, $2$). For convenience, we assume
$\alpha$ is a real number throughout this paper and $|\alpha\rangle$
$=$ $\sum_{n=0}^{\infty}A_{n}|n\rangle$ where
$A_{n}=\alpha^{n}e^{-\alpha^2/2}/\sqrt{n!}$. In the picture rotating
at local field frequency, the interaction Hamiltonian under the
rotating-wave approximation is ($\hbar$ $=$ $1$):
\begin{eqnarray}\label{e1}
&H_{I}=\sum_{i=1}^{2}\bigg[\Delta
a_{i}^{\dagger}a_{i}+g\big(S_{i}^{+}a_{i}+S_{i}^{-}a_{i}^{\dagger}\big)\bigg]
+J\big(a_{1}^{\dag}a_{2}+a_{1}a_{2}^{\dag}\big),&
\end{eqnarray}
where $S_{i}^{+}$ $=$ $|e_{i}\rangle\langle g_{i}|$ and $S_{i}^{-}$
$=$ $|g_{i}\rangle\langle e_{i}|$ with $|e_{i}\rangle$ and
$|g_{i}\rangle$ being the excited state and ground state of the
$i$th qubit. $a_{i}^{\dag}$ and $a_{i}$ are respectively the
creation and annihilation operators for the $i$th field mode, $g$
describes the coupling strength between the qubit and field mode,
and $\Delta$ is the detuning between the qubit's transition and
field mode. $J$ represents the coherent photon hopping strength
between two resonators. In the limit of zero hopping ($J=0$) and
zero detuning ($\Delta=0$), the present system reduces to a pair of
noninteracting atom-cavity systems\cite{PRL-96-080501-2006}, each
described by the resonant JC interaction.

We assume the initial state of two qubits is a pure state
$|\psi_{a}(0)\rangle$, then the evolution of the qubit-field system
is:
\begin{eqnarray}\label{e2}
|\Psi(t)\rangle =
e^{-iH_{I}t}|\psi_{a}(0)\rangle|\alpha\rangle_{1}|\alpha\rangle_{2}.
\end{eqnarray}
Since the total excitation number operator $\hat{M}$ $=$
$\sum_{i=1}^{2}(|e_{i}\rangle\langle e_{i}|+a_{i}^{+}a_{i})$
commutes with $H_{I}$, the excitation number of the qubit-field
system is conserved during its evolution. For specific $\alpha$, we
truncate the total excitation number at $M$ in the Hilbert space,
therefore the system evolution is :
\begin{eqnarray}\label{e3}
|\Psi(t)\rangle &=&\ \sum_{N=1}^{M}
 \sum^{N-1}_{\mbox{\tiny$\begin{array}{c}
n_{1}=0\\
n_{2}=N-1-n_{1}\end{array}$}}U_{egn_{1}n_{2}}^{N}(t)|e_{1}g_{2}\rangle|n_{1}\rangle_{1}|n_{2}\rangle_{2}\cr&&
+\sum_{N=1}^{M}\sum^{N-1}_{\mbox{\tiny$\begin{array}{c}
n_{1}=0\\
n_{2}=N-1-n_{1}\end{array}$}}U_{gen_{1}n_{2}}^{N}(t)|g_{1}e_{2}\rangle|n_{1}\rangle_{1}|n_{2}\rangle_{2}\cr&&
+\sum_{N=0}^{M}\sum^{N}_{\mbox{\tiny$\begin{array}{c}
n_{1}=0\\
n_{2}=N-n_{1}\end{array}$}}U_{ggn_{1}n_{2}}^{N}(t)|g_{1}g_{2}\rangle|n_{1}\rangle_{1}|n_{2}\rangle_{2}\cr&&
+\sum_{N=2}^{M}\sum^{N-2}_{\mbox{\tiny$\begin{array}{c}
n_{1}=0\\
n_{2}=N-2-n_{1}\end{array}$}}U_{een_{1}n_{2}}^{N-2}(t)|e_{1}e_{2}\rangle|n_{1}\rangle_{1}|n_{2}\rangle_{2},
\end{eqnarray}
where $U_{egn_{1}n_{2}}^{N}(t)$, $U_{gen_{1}n_{2}}^{N}(t)$,
$U_{ggn_{1}n_{2}}^{N}(t)$ and $U_{een_{1}n_{2}}^{N-2}(t)$ are the
coefficients of the corresponding state components
$|e_{1}g_{2}\rangle|n_{1}\rangle_{1}|n_{2}\rangle_{2}$,
$|g_{1}e_{2}\rangle|n_{1}\rangle_{1}|n_{2}\rangle_{2}$,
$|g_{1}g_{2}\rangle|n_{1}\rangle_{1}|n_{2}\rangle_{2}$ and
$|e_{1}e_{2}\rangle|n_{1}\rangle_{1}|n_{2}\rangle_{2}$,
respectively.

\section{Qubit-qubit entanglement}

We assume each field is initially prepared in the coherent state
with mean photon number $\bar{n}=\alpha^{2}$, and discuss the
entanglement dynamics of two qubits with different initial states
based on numerical truncation. For example, when $\bar{n}=1$, the
Hilbert space is safely cut off at $M=15$; when $\bar{n}=100$, the
Hilbert space is safely cut off at $M=210$. When $\alpha\gg1$, the
width of the photon number distribution obeys $1\ll\Delta
n\ll\alpha^2$. Therefore it is safely to truncate the Fock state
basis to $M=10+2\alpha^2$ for special $\alpha$ value.

In this section, we use the Wootters's concurrence $C$ as the
entanglement measure for two qubits expressed in the standard qubit
basis $\{|e_{1}e_{2}\rangle$, $|e_{1}g_{2}\rangle$,
$|g_{1}e_{2}\rangle$, $|g_{1}g_{2}\rangle\}$, which is defined as
\cite{PRL-1998-80-2245}:
\begin{eqnarray}\label{e4}
C=\max\bigg\{ 0,
\sqrt{\lambda_{1}}-\sqrt{\lambda_{2}}-\sqrt{\lambda_{3}}-\sqrt{\lambda_{4}}
\bigg\},
\end{eqnarray}
where $\lambda_{1}$, $\lambda_{2}$, $\lambda_{3}$ and $\lambda_{4}$
are the eigenvalues arranged in decreasing order of the following
matrix:
\begin{eqnarray}\label{e5}
\xi&=&\rho(\sigma_{y}\otimes\sigma_{y})\rho^{*}(\sigma_{y}\otimes\sigma_{y}),
\end{eqnarray}
where $\sigma_{y}$ is the corresponding Pauli matrix, and $\rho$ is
the two-qubit reduced density matrix. $C=0$ is for two unentangled
qubits, and $C=1$ stands for the maximal entanglement of two qubits.
Thus, the reduced density matrix $\rho$ for the qubits is calculated
by tracing out the fields,
and the elements of $\rho$ are straightforwardly solved by the
numerical simulation.

\begin{figure}[H]\label{fig2}
\centering
\includegraphics[width=0.8\columnwidth]{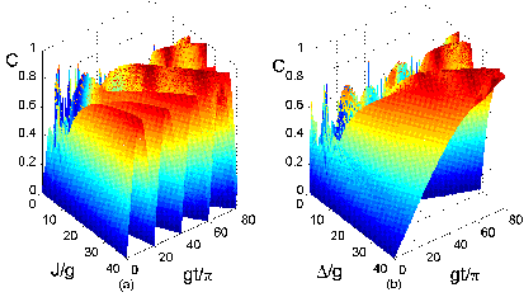}
\caption{(Color online) The concurrence $C$ for two qubits depends
on the interaction time $gt/\pi$ and: (a) the hopping strength $J/g$
with $\Delta=5g$; (b) the detuning $\Delta/g$ with $J=5g$, where
$\alpha=1$ and the qubits are initially in the state
$|e_{1}g_{2}\rangle$.}
\end{figure}

\begin{figure}[H]\label{fig3}
\centering
\includegraphics[width=0.8\columnwidth]{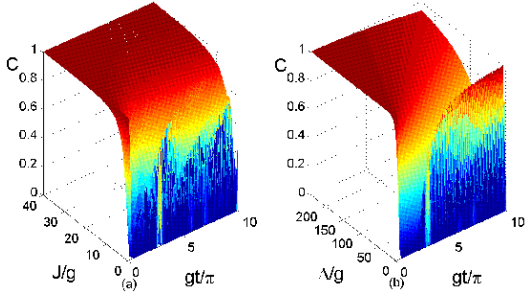}\caption{(Color online) The
concurrence $C$ for two qubits depends on the interaction time
$gt/\pi$ and: (a) the hopping strength $J/g$ with $\Delta=0$; (b)
the detuning $\Delta/g$ with $J=0$, where $\alpha=1$ and the qubits
are initially in the maximally entangled state
$(|e_{1}g_{2}\rangle+|g_{1}e_{2}\rangle)/\sqrt{2}$.}
\end{figure}

We plot the concurrence of two qubits with initial qubit state
$|\psi_{a}(0)\rangle$ $=$ $|e_{1} g_{2}\rangle$ against the
interaction time $gt$ and the hopping strength $J$ (the detuning
$\Delta$) in Fig. 2(a) (Fig. 2(b)). The appearance of $C=1$ is
clearly visible in Fig. 2 (a) and (b). This is because when large
detuning condition is satisfied, the probability for energy exchange
between the qubits and field modes is close to zero, and the two
qubits couple with each other by exchanging the virtual excitation
of field modes. This can be understood by the effective Hamiltonian
\cite{PRA-78-063805-2008,PRL-85-2392-2000,JOSAB-29-2379-2012} :
\begin{eqnarray}\label{e6}
H_{eff}&=&-\sum_{i=1}^{2}\big[(\frac{g^2}{2\Delta_{1}^{'}}b_{1}^{\dagger}b_{1}+\frac{g^2}{2\Delta_{2}^{'}}b_{2}^{\dagger}b_{2})
(|e_{i}\rangle\langle e_{i}|-|g_{i}\rangle\langle g_{i}|)\cr&&
+(\frac{g^2}{2\Delta_{1}^{'}}+\frac{g^2}{2\Delta_{2}^{'}}
)|e_{i}\rangle\langle e_{i}|\big]+\lambda(
S_{1}^{+}S_{2}^{-}+S_{1}^{-}S_{2}^{+}),
\end{eqnarray}
where $b_{1}=(a_{1}+a_{2})/\sqrt{2}$,
$b_{2}=(a_{1}-a_{2})/\sqrt{2}$, $\Delta_{1}^{'}=\Delta+J$,
$\Delta_{2}^{'}=\Delta-J$, and
$\lambda=\frac{g^2}{2\Delta_{1}^{'}}-\frac{g^2}{2\Delta_{2}^{'}}$.
Under the large detuning condition $\Delta_{1}^{'}$,
$\Delta_{2}^{'}$ $\gg$ $\sqrt{\bar{n}+1}g/\sqrt{2}$, the evolution
of the two-qubit system is $|\Psi(t)\rangle_{eff}$ $=$ $e^{-i\lambda
t}$ $[\cos(\lambda t)|e_{1}g_{2}\rangle$ $-$ $i\sin(\lambda
t)|g_{1}e_{2}\rangle]$, which is independent of the field states and
accounts for the periodic oscillation behavior in Fig. 2. Therefore,
for large hopping strength or local qubit-field detuning, the
coupled CV systems can generate the maximal qubit-qubit
entanglement.

We now consider the case that are initially prepared in the maximally entangled state 
$|\psi_{a}(0)\rangle$ $=$
$(|e_{1}g_{2}\rangle+|g_{1}e_{2}\rangle)/\sqrt{2}$. For $J\gg g$ in
Fig. 3(a) or $\Delta\gg g$ in Fig. 3(b), the maximal qubit-qubit
entanglement can be frozen. This is because that the large detuning
condition $\Delta_{1}^{'}$, $\Delta_{2}^{'}$ $\gg$
$\sqrt{\bar{n}+1}g/\sqrt{2}$ is satisfied in this case, therefore,
the effective Hamiltonian of Eq. (\ref{e6}) becomes valid in the
system evolution, in which the maximally entangled state for two
qubits becomes an eigenstate of this effective Hamiltonian.
\begin{figure}\label{fig4}
 \label{Fig.sub.b}
\includegraphics[width=1\columnwidth]{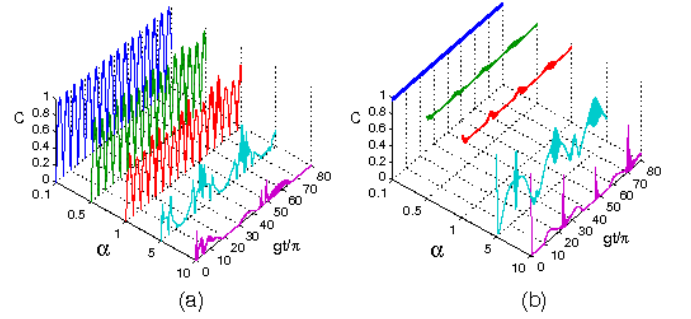}
\caption{(Color online) The concurrence $C$ for two qubits versus
the interaction time $gt/\pi$ for different amplitudes $\alpha$ of
the initial coherent state with $\Delta=0$ and $J=10g$. The qubits
are initially in the state: (a) $|e_{1}g_{2}\rangle$; (b)
$(|e_{1}g_{2}\rangle+g_{1}e_{2}\rangle)/\sqrt{2}$. }
\end{figure}
\begin{figure}\label{fig5}
\centering
\includegraphics[width=1\columnwidth]{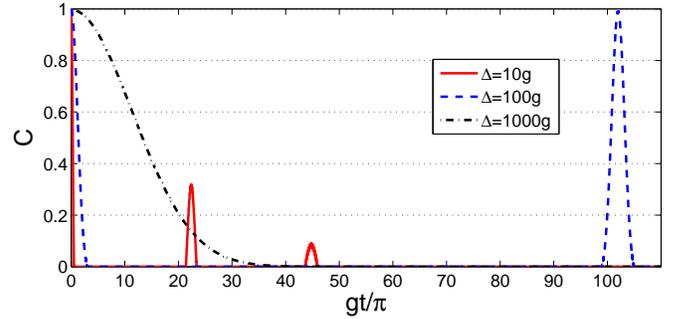} \caption{(Color online)
The concurrence $C$ for two qubits versus the interaction time
$gt/\pi$ with $\alpha=10$ and $J=0$. The qubits are initially in the
state $(|e_{1}g_{2}\rangle+g_{1}e_{2}\rangle)/\sqrt{2}$.}
\end{figure}
We also plot the two-qubit concurrence against the interaction time
for different amplitudes of initial coherent state in Fig. 4. The
result shows that the concurrence $C$ becomes less stable as the
amplitude $\alpha$ increases for fixed $J$ and $\Delta$ values. This
is due to the fact that the probability that the qubits exchange
energy with the fields increases with the mean photon numbers
$\bar{n}$. When the mean photon numbers are large enough, the large
detuning condition $\Delta_{1}^{'}$, $\Delta_{2}^{'}$ $\gg$
$\sqrt{\bar{n}+1}g/\sqrt{2}$ is not satisfied, therefore the
effective Hamiltonian of Eq. (\ref{e6}) loses its validness in the
system evolution. Unlike the resonant situation in Ref.
\cite{OL-33-270-2008}, the effects of the detuning $\Delta$ on the
two-qubit concurrence $C$ are considered in Fig. 5. We observe that
the period of the collapse and revival of qubit-qubit entanglement
is delayed when the detuning $\Delta$ increases. Especially, even
for large mean photon number $\bar{n}=100$, the entanglement can be
fully revival if the the detuning is large enough.

\begin{figure}\label{fig6}
\centering
\includegraphics[width=1\columnwidth]{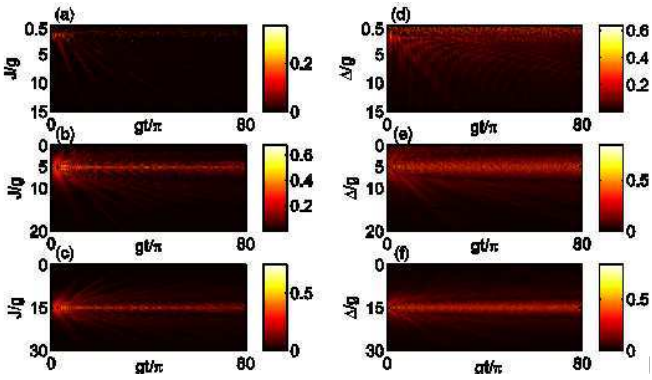}
\caption{(Color online) The concurrence $C$ for two qubits depends
on the interaction time $gt/\pi$ and the hopping strength $J/g$ when
the qubits are initially in the state $|e_{1}e_{2}\rangle$ with
$\alpha=1$: (a) $\Delta=0.5g$; (b) $\Delta=5g$; (b) $\Delta=15g$.
The concurrence $C$ for two qubits depends on the interaction time
$gt/\pi$ and the detuning $\Delta/g$ when the qubits are initially
in the state $|g_{1}g_{2}\rangle$ with $\alpha=1$: (d) $J=0.5g$; (e)
$J=5g$; (f) $J=15g$.}
\end{figure}

When the qubits are both initially in their excited states
$|e_{1}e_{2}\rangle$ or ground states $|g_{1}g_{2}\rangle$, we find
that the qubit-qubit entanglement could be induced by two initially
coherent state fields with appropriate parameters in Fig. 6. The
most obvious feature of the two-qubit concurrence in Fig. 6 is that
the birth and death of qubit-qubit entanglement only appear in the
regime when $\Delta$ approaches $J$, and the peaks of the two-qubit
concurrence do not decrease monotonically as the interaction time
increases. For $\Delta \sim J$, the qubit transition is nearly
resonant with one delocalized modes which mediates the qubit-qubit
entanglement. The qubits are hardly affected by the other
nonresonant mode, and undergo relatively fast Rabi oscillations.
While the subsystem between two qubits and the other delocalized
mode experiences relatively slow Rabi oscillations, which
essentially induces the qubit-qubit entanglement.

\section{Photon hopping and detuning modulations for entanglement reciprocation}

In this section, we study the effects caused by photon hopping and
qubit-field detuning on the process of entanglement reciprocation
followed in a recent proposal \cite{PRL-96-080501-2006}.

We consider that two qubits are initially in the maximally entangled
state
\begin{eqnarray}\label{er7}
|\psi_{a}(0)\rangle=\frac{1}{\sqrt{2}}(|e_{1}g_{2}\rangle+|g_{1}e_{2}\rangle),
\end{eqnarray}
and two resonators are initially prepared in their coherent state
fields $|\psi_{f}(0)\rangle=|\alpha\rangle_{1}|\alpha\rangle_{2}$.
After the interaction time $t$, the qubit-field system evolves to
the state
$|\Psi_{af}(t)\rangle=e^{-iH_{I}t}|\psi_{a}(0)\rangle|\psi_{f}(0)\rangle$.
Similar to Ref. \cite{PRL-96-080501-2006}, we postselect the fields
conditioned on two qubits leaving their resonators both in the
ground states. The fields after this postselection are in a pure
state:
\begin{eqnarray}\label{er8}
|\Psi_f(t)\rangle=N_{f}\sum_{l=0}^{M}\sum_{m=0}^{M}U_{l,m}(t)|l\rangle_{1}|m\rangle_{2},
\end{eqnarray}
where $|l\rangle_1$ and $|m\rangle_2$ represent the Fock state basis
for the field 1 and 2, respectively. $U_{l,m}(t)$ is the
time-dependent coefficient of state component
$|l\rangle_{1}|m\rangle_{2}$, and $N_{f}$ is the normalization
constant.

We take von Neumann entropy to measure field-field entanglement.
Taking a partial trace over the field 2, we can obtain the reduced
density matrix for the field 1:
\begin{eqnarray}\label{er9}
\rho_{f1}&=&
Tr_{2}\bigg(|\Psi_f(t)\rangle\langle\Psi_f(t)|\bigg)\cr&&
=\sum_{m=0}^{M}\sum_{l=0}^{M}
\sum_{l^{'}=0}^{M}N_{f}N_{f}^{*}U_{l,m}(t)U_{l^{'},m}^{*}(t)|l\rangle_1\langle
l^{'}|,
\end{eqnarray}
and the von Neumann entropy of the field 1 is explicitly calculated
by $\epsilon=-Tr\bigg(\rho_{f1}\log_{2}\rho_{f1}\bigg)$.

\subsection{Photon hopping modulation}

With different photon hopping strengths between the resonators, the
evolution dynamics for the qubit-field system in the present paper
is very different from that in Ref. \cite{PRL-96-080501-2006}, and
four subsystems involving the Jaynes-Cummings interaction and the
field-field interaction constitute the whole system.

We plot the entanglement $\varepsilon$ of the field against the
amplitude $\alpha$ and the interaction time $gt$ in Fig. 7 [(a) -
(c)]. For $\alpha=0$, the qubit-field system is simplified to the
model with two coupled vacuum fields in Ref.
\cite{PRA-78-063805-2008}, where $\varepsilon$ can be 1 for sure.
Fig. 7 [(d) - (f)] show that the probability $P$ of two qubits
leaving the resonators in their ground states, where $P$ $=$
$\sum_{l=0}^{M}\sum_{m=0}^{M}|U_{l,m}(t)|^2$. It is interesting to
see that when $\alpha<1$, $\varepsilon$ can be larger than $1$ and
$\varepsilon$ becomes larger as $\alpha$ increases, which is very
different from the result with the maximal field-field entanglement
$\varepsilon=1$ in Ref. \cite{PRL-96-080501-2006}. This is because
that the entanglement between the two Jaynes-Cummings subsystems is
not conserved due to their interaction, which can help to improve
the entanglement between the fields. However, it should be noted
that the photon-hopping itself can not lead to the field-field
entanglement for the initial coherent states. The Jaynes-Cummings
interaction makes each field deviate from the coherent state so that
the photon hopping can enhance the field-field entanglement. When
$\alpha>0$, the oscillation behaviors of the probability $P$ in Fig.
7 (d) and (e) are similar to that in Ref. \cite{PRL-96-080501-2006}.
When the large detuning condition is satisfied, as the case with
$J\gg \sqrt{\alpha^2+1}g/\sqrt{2}$ plotted in Fig. 7 (c) and (f),
the probability for energy exchange between the qubits and field
modes is close to zero, which makes the probability $P$ in Fig. 7
(f) experience no oscillation behaviors for small $\alpha$. It is
interesting to observe that when $\alpha>1$, the field-field
entanglement $\varepsilon$ keeps larger than $1$ whenever the qubits
leave the resonators in their ground states after the first moments
of oscillation.

\begin{figure}
\centering  \label{fig7}
\includegraphics[width=0.9\columnwidth]{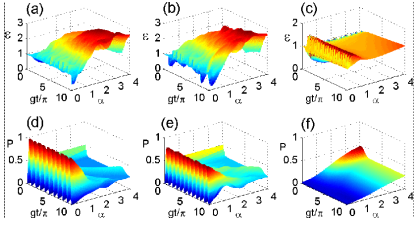}
 \caption{(Color online) The degree of entanglement $\varepsilon$ for the field
 1 depends on the amplitude $\alpha$ of the initial coherent state and the
 interaction time $gt$ (in units of $\pi$) when $\Delta=0$: (a) $J=0.1g$; (b) $J=g$; (c) $J=10g$.
 Probability $P$ for the qubits leaving the resonators in their
ground states when $\Delta=0$: (d) $J=0.1g$; (e) $J=g$; (f) $J=10g$.
}
\end{figure}

\begin{figure}
\centering  \label{fig8}
\includegraphics[width=0.9\columnwidth]{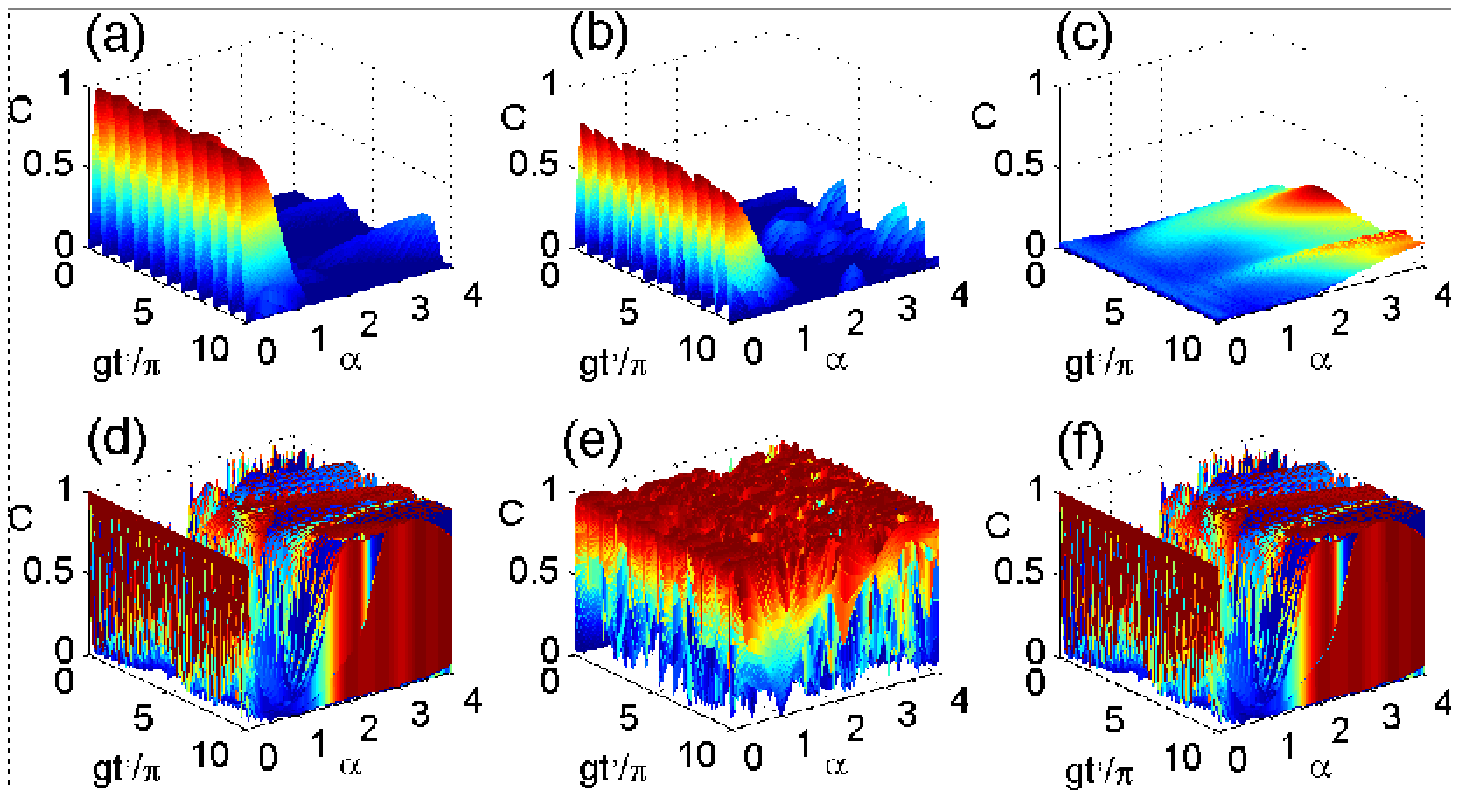}
 \caption{(Color online) The concurrence $C$ for the second
pair of qubits versus on the amplitude $\alpha$ of the initial
coherent state and the interaction time $gt^{'}$ ($gt^{'}=gt$) when
$\Delta=0$: (a) $J=0.1g$; (b) $J=g$; (c) $J=10g$. The concurrence
for the second pair of qubits after postselecting the cavity fields
both in their coherent states when $\Delta=0$: (d) $J=0.1g$; (e)
$J=g$; (f) $J=10g$.}
\end{figure}

\begin{figure}
\centering \label{fig9}
\includegraphics[width=0.9\columnwidth]{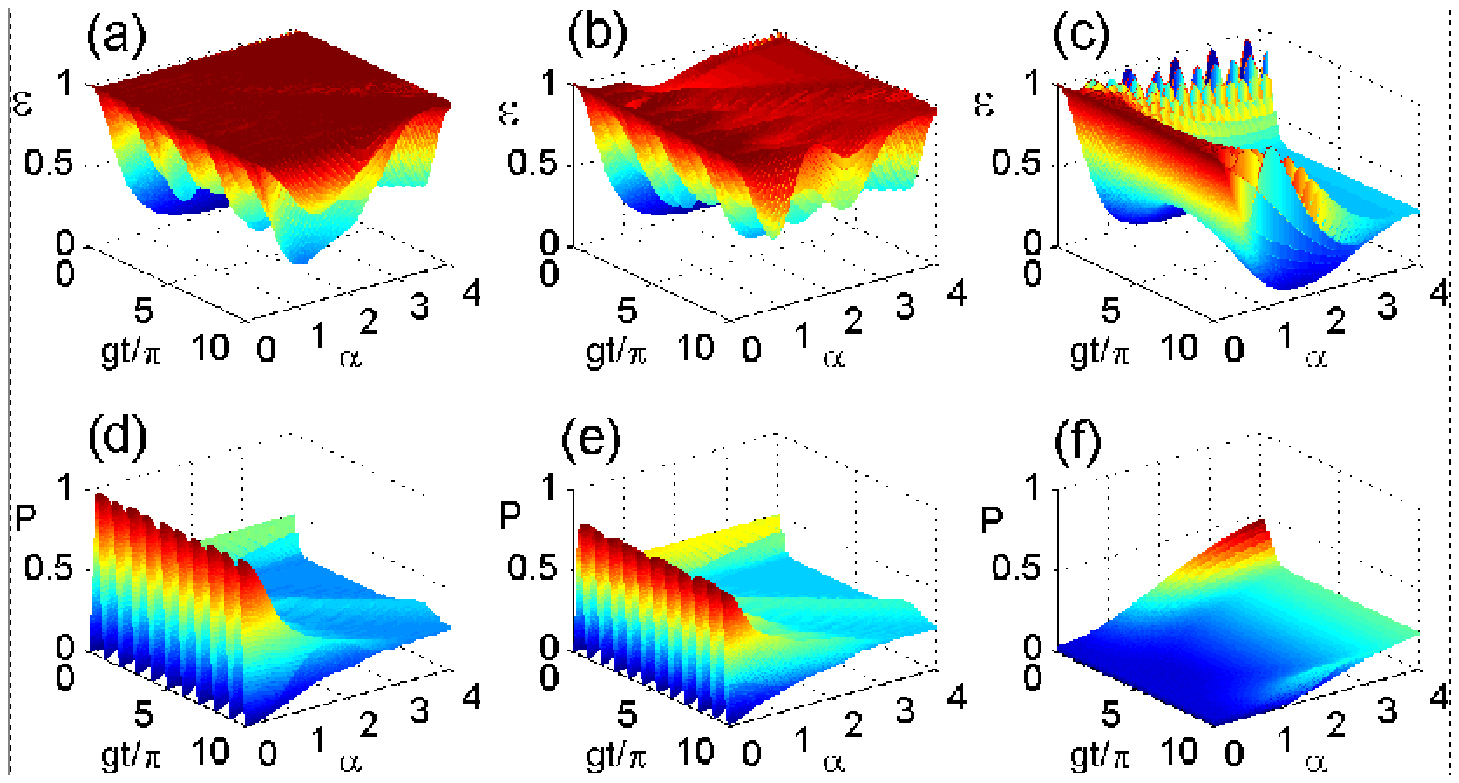}
 \caption{(Color online) The degree of entanglement $\varepsilon$ for the field
 1 depends on the amplitude $\alpha$ of the initial coherent state and the
 interaction time $gt$ (in units of $\pi$) when $J=0$: (a) $\Delta=0.1g$; (b) $\Delta=g$; (c) $\Delta=10g$.
 Probability $P$ for the qubits leaving the resonators in their
ground states when $J=0$: (d) $\Delta=0.1g$; (e) $\Delta=g$; (f)
$\Delta=10g$.}
\end{figure}

\begin{figure}
\centering  \label{fig10}
\includegraphics[width=0.9\columnwidth]{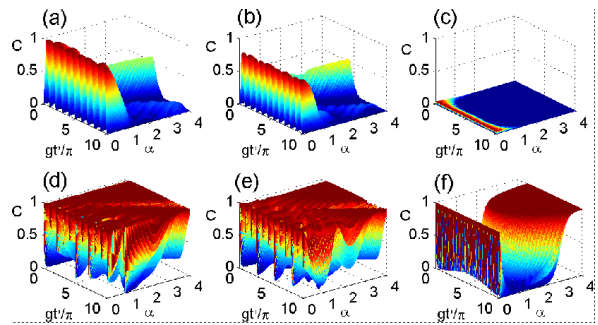}
 \caption{(Color online) The concurrence $C$ for the second
pair of qubits versus the amplitude $\alpha$ of the initial coherent
state and the interaction time $gt^{'}$ ($gt^{'}=gt$) when $J=0$:
(a) $\Delta=0.1g$; (b) $\Delta=g$; (c) $\Delta=10g$. The concurrence
for the second pair of qubits after postselecting the fields both in
their coherent states when $\Delta=0$: (d) $\Delta=0.1g$; (e)
$\Delta=g$; (f) $\Delta=10g$.}
\end{figure}
To see how a pair of unentangled qubits interact with the
highly-entangled coupled CV fields, we send the second pair of
qubits with the initial state $|g_{1}g_{2}\rangle$ into the
respective resonator, containing the rest entangled field state
$|\Psi_{f}(t)\rangle$. Applying the Hamiltonian in Eq. (\ref{e1})
again to obtain the system evolution after an interaction time
$t^{'}$:
\begin{eqnarray}\label{er10}
|\Psi_{af}^{'}(t^{'})\rangle=e^{-iH_{I}t^{'}}|g_{1}g_{2}\rangle|\Psi_{f}(t)\rangle.
\end{eqnarray}
To investigate the entanglement dynamics of the second pair of
qubits, we trace $|\Psi_{af}^{'}(t^{'})\rangle$ over the field
variables and obtain the reduced density matrix $\rho^{'}$ for two
qubits:
\begin{eqnarray}\label{er11}
\rho^{'}=\sum_{l=0}^{M}\sum_{m=0}^{M}\ _{1}\langle l|_{2}\langle
m|\bigg(|\Psi_{af}^{'}(t^{'})\rangle\langle\Psi_{af}^{'}(t^{'})|\bigg)|m\rangle_{2}|l\rangle_{1}.
\end{eqnarray}
Taking the definition of Eq. (\ref{e4}), we plot the concurrence $C$
against the amplitude $\alpha$ and the interaction time $gt^{'}$ in
Fig. 8 [(a) - (c)]. We find that if the photon hopping strength is
small enough, as shown in Fig. 8 (a), two qubits are able to become
maximally entangled when interacting with highly-entangled field
states. However, two qubits are not able to become maximally
entangled as the photon hopping strength increases even for
$\alpha=0$, as shown in Fig. 8 (b) and (c), due to the entanglement
loss induced by the photon hopping.

In order to improve the degree of entanglement for the second pair
of qubits, we measure the fields with the projection onto the
coherent state of amplitude $\alpha$ similar to Ref.
\cite{PRL-96-080501-2006}:
\begin{eqnarray}\label{er12}
|\Psi_{a}(t^{'})\rangle&=&_{2}\langle\alpha|_{1}\langle\alpha|\otimes|\Psi_{af}^{'}(t^{'})\rangle.
\end{eqnarray}
The concurrence of the qubits after measuring the fields is plotted
in Fig. 8 [(d) - (f)]. After postselecting the cavity fields both in
their coherent states, the entanglement of two qubits exhibits sharp
oscillating behaviors and can be $1$.

\subsection{Detuning modulation}

Assume the photon-hopping becomes so weak that can be ignored, we
directly generalize the entanglement reciprocation with resonant
Jaynes-Cummings interaction in Ref. \cite{PRL-96-080501-2006} to the
situation with detune Jaynes-Cummings interaction.

Based on the processes from Eq. (\ref{er7}) to Eq. (\ref{er9}), we
plot the entanglement $\varepsilon$ of the field against the
amplitude $\alpha$ and the interaction time $gt$ in Fig. 9 [(a) -
(c)], and the probability $P$ of two qubits leaving the resonators
in their ground states in Fig. 9 [(d) - (f)] under different
qubit-field detunings. For $\alpha=0$, $\varepsilon$ can be 1 for
sure, meaning the complete ebit in two qubits can be transferred to
the fields under different detunings. Even when the detuning
increases to a certain degree, such as $\Delta=g$ plotted in Fig. 9
(b) and (e), the fields can be with one complete ebit whenever the
qubits leave their resonators, which implies the entanglement
reciprocation \cite{PRL-96-080501-2006} tolerates the qubit-field
detuning within a wide range. However, for large detuning case
$\Delta\gg \sqrt{\alpha^2+1}g/\sqrt{2}$ in Fig. 9 (c) and (f),
$\varepsilon$ is not able to keep in $1$, which is very different
from that in the resonant situation \cite{PRL-96-080501-2006}. This
is because under the large detuning regime, the maximally entangled
state for two qubits is an eigenstate of the effective Hamiltonian
in Eq. (\ref{e6}), therefore an ebit for the qubits can not be
completely transferred to the fields and the probability for finding
the qubits both in their ground states is rather low.

In order to see whether it is possible for two unentangled qubits to
retrieve the ebit from rest entangled CV systems under different
qubit-field detunings. We send the second pair of qubits with the
initial state $|g_1g_2\rangle$ into their respective resonators,
which are in the state $|\Psi_{f}(t)\rangle$. Based on the equations
from Eq. (\ref{er10}) to Eq. (\ref{er12}), we plot the two-qubit
concurrence against the amplitude of the initial coherent state and
the interaction time $gt^{'}$ in Fig. 10 [(a) - (c)], and the
two-qubit concurrence after measuring the fields with the projection
onto their coherent states of amplitude $\alpha$ in Fig. 10 [(d) -
(f)] under different qubit-field detunings. We find that if the
detuning is small enough, as shown in Fig. 10 (a), two qubits are
able to retrieve an ebit from the entangled CV systems for
$\alpha=0$. As the detuning increases, the entanglement that the
qubits can retrieve from the entangled CV systems decreases,
especially for the large detuning situation in Fig. 10 (c), the
entanglement retrieved by the qubits is close to zero. This is
because the probability for energy exchange between the qubits and
CV systems is close to zero under the condition $\Delta\gg
\sqrt{\alpha^2+1}g/2$. After measuring the fields, it is interesting
to observe that two qubits can become the maximally entangled state,
as plotted in Fig. 10 [(d) - (f)]. Even for the large detuning
situation plotted in Fig. 10 (f), the qubits can become the
maximally entangled when $\alpha$ is large enough.

Therefore, the entanglement dynamics between the qubits and the
coupled CV systems here becomes different from that in the system
involving two uncoupled CV systems. When there is photon hopping
strength between two CV systems, the field-field entanglement can be
larger than 1 and increases as the initial amplitude of the coherent
state grows. By postselecting the fields in their coherent states,
the entanglement of two initially unentangled qubits can be largely
improved.

\section{Conclusion}

To conclude, we have numerically study the exact dynamics for two
qubits based on the coupled coherent state fields, and the
modulations on the entanglement reciprocation induced by photon
hopping between two CV systems and the qubit-field detuning. In far
off-resonant regime, the maximal entanglement of two qubits can be
generated with the initial qubit state in which one qubit is in the
excited state and the other is in the ground state, while the
initially maximal two-qubit entanglement can be frozen and fully
revival even for large mean photon number. For example, the maximal
entanglement is fully revival even for large mean photon number
$\bar{n}=100$ by choosing $\Delta=100g$, and the period of the
entanglement collapse and revival is delayed as the detuning
increases further. When the qubits are both initially in their
excited states or ground states, the qubit-qubit entanglement
concentrates to appear at $\triangle \sim J$, and its peaks do not
decrease monotonically as the interaction time increases. When there
is photon hopping strength between two CV systems, the field-field
entanglement can be larger than 1 and increases as the initial
amplitude of the coherent state grows. By postselecting the fields
both in their coherent states, the entanglement of the qubits can be
largely improved. Even when the detuning increases to a certain
degree, such as $\Delta=g$, the fields can be with one complete ebit
whenever the qubits leave their cavities. Our present idea provides
the fundamental setup for manipulating the qubit-qubit entanglement
with coupled CV systems and is applicable for different physical
systems.

\section{Acknowledgement}

This work is supported by the Major State Basic Research Development
Program of China under Grant No. 2012CB921601, the National Natural
Science Foundation of China under Grant No. 11374054, No. 11305037,
No. 11347114, and No. 11247283, the Natural Science Foundation of
Fujian Province under Grant No. 2013J01012, and  the funds from
Fuzhou University under Grant No. 022513, Grant No. 022408, and
Grant No. 600891.


\begin{thebibliography}{999}
\bibitem{RMP-70-101-1998} M.~B.~Plenio and P.~L.~Knight, Rev. Mod. Phys. 70, 101 (1998).
\bibitem{PRL-92-127902-2004} J.~M.~Raimond, M.~Brune, and S.~Haroche, Rev. Mod. Phys. 73, 565 (2001).
\bibitem{PRL-91-070402-2003} F.~Benatti, R.~Floreanini, and M.~Piani, Phys. Rev. Lett. 91, 070402 (2003).
\bibitem{PRL-85-2392-2000} S.~B.~Zheng and G.~C.~Guo, Phys. Rev. Lett. 85, 2392 (2000).
\bibitem{PRL-87-037902-2001} S.~Osnaghi, P.~Bertet, A.~Auffeves,
P.~Maioli, M.~Brune, J.~M.~Raimond, and S.~Haroche, Phys. Rev. Lett.
87, 037902 (2001).

\bibitem{OC-176-265-2000} S.~B.~Zheng, Opt. Commun. 176, 265
(2000).
\bibitem{Nature-411-166-2001} P.~Bertet, \emph{et al.}, Nature 411,
166 (2001).
\bibitem{Nature-453-1023-2008} H.~J.~Kimble, Nature 453, 1023
(2008).
\bibitem{QuantumOptics} M.~O.~Scully and M.~S.~Zubairy, Quantum Optics (Cambridge University
Press) (1997).
%
\bibitem{RMP-77-513-2005} S.~L.~Braunstein and P.~Van.~Loock, Rev. Mod.
Phys. 77, 513 (2005); and the references therein.
\bibitem{JPA-40-7821-2007} G.~Adesso and F.~Illuminati, J. Phys. A: Math. Theor. 40, 7821 (2007); and the references therein.


\bibitem{PRA-75-032336-2007} C.~Federico, L.~Alfredo, and G.~A.~P.~Matteo, Phys. Rev. A 75, 032336 (2007).
\bibitem{JPB-44-105501-2011} P.~Blanco and D.~Mundarain, J. Phys. B: At. Mol. Opt.
Phys. 44, 105501 (2011).

\bibitem{PRA-79-062317-2009} D.~Ballester, Phys. Rev. A 79, 062317 (2009).
\bibitem{JPB-39-5143-2006} L.~Zhou and G.~H.~Yang, J. Phys. B: At. Mol. Opt. Phys. 39,
5143 (2006).

\bibitem{PRL-96-080501-2006} J.~Lee, M.~Paternostro, M.~S.~Kim, and S.~Bose, Phys. Rev. Lett. 96,
080501 (2006).
\bibitem{PRL-98-140504-2007} M.~Paternostro, M.~S.~Kim, and G.~M.~Palma, Phys. Rev. Lett. 98,
140504 (2007).


\bibitem{OL-33-270-2008} M.~Y\"{o}na\c{c} and J.~H.~Eberly, Opt. Lett. 33, 270 (2008);
M.~Y\"{o}na\c{c} and J.~H.~Eberly, Phys. Rev. A 82, 022321 (2010).

\bibitem{PRA-86-052315-2012} Y.~Q.~Guo, J.~Li, T.~C.~Zhang, and M.~Paternostro, Phys. Rev. A 86, 052315 (2012).

\bibitem{PRA-78-063805-2008} C.~D.~Ogden, E.~K.~Irish, and M.~S.~Kim, Phys. Rev. A 78, 063805 (2008).
\bibitem{LPR-2-527-2008}     M.~J.~Hartmann, F.~G.~S.~L.~Branda\~{o}, and M.~B.~Plenio, Laser Photon. Rev. 2, 527 (2008).
\bibitem{PRL-96-010503-2006} A.~Serafini, S.~Mancini, and S.~Bose, Phys. Rev. Lett. 96, 010503
(2006); Z.~Q.~Yin and F.~L.~Li, Phys. Rev. A 75, 012324 (2007).

\bibitem{PRA-82-042327-2010} S.~B.~Zheng, C.~P.~Yang, and F.~Nori, Phys. Rev. A 82, 042327 (2010).
\bibitem{JOSAB-29-2379-2012} L.~T.~Shen, Z.~B.~Yang, H.~Z.~Wu, X.~Y.~Chen, and S.~B.~Zheng, J. Opt. Soc. Am. B 29, 2379 (2012).
\bibitem{Nature-484-195-2012} S.~Ritter, C.~N\"{o}lleke, C.~Hahn, A.~Reiserer, A.~Neuzner, M.~Uphoff, M.~M\"{u}cke, E.~Figueroa, J.~Bochmann, and G.~Rempe, Nature 484, 195 (2012).

\bibitem{PRL-1998-80-2245} W.~K.~Wootters, Phys. Rev. Lett. 80, 2245 (1998).


%











\end{thebibliography}
\end{document}